\documentclass[runningheads]{svmult}

\usepackage{graphicx}  % standard LaTeX graphics tool for including eps-figure files

\usepackage{physprbb}  % modified textarea for proceedings, lecture notes, and the like.

\def\lsim{\hbox{ \rlap{\raise 0.425ex\hbox{$<$}}\lower 0.65ex\hbox{$\sim$} }}
\def\gsim{\hbox{ \rlap{\raise 0.425ex\hbox{$>$}}\lower 0.65ex\hbox{$\sim$} }}
\def\etal{{\it et al. }}

\def\pp{\noindent\parshape 2 0.0 truecm 12.1 truecm 0.6 truecm 11.5 truecm}

\begin{document}

\title*{Challenges for Cluster Analysis in a Virtual Observatory
\footnote{ To appear in: {\sl Statistical Challenges in Modern Astronomy III},
eds. E. Feigelson and G.J. Babu, chapter 13, p. 125, New York: Springer Verlag
(2002).}
}
\toctitle{Cluster Analysis in a Virtual Observatory}
\titlerunning{Cluster Analysis in a Virtual Observatory}
\author{S.G. Djorgovski \inst{1}
\and    R. Brunner \inst{1}
\and    A. Mahabal \inst{1}
\and    R. Williams \inst{2}
\and    R. Granat \inst{3}
\and    P. Stolorz \inst{3}
}
\authorrunning{Djorgovski et al.}
\institute{Palomar Observatory, Caltech, Pasadena, CA 91125, USA
\and Center for Advanced Computing Research, Caltech, Pasadena, CA 91125, USA
\and Jet Propulsion Laboratory, Pasadena, CA 91109, USA
}

\maketitle              % typesets the title of the contribution

\begin{abstract}
here has been an unprecedented and continuing growth in the
volume, quality, and complexity of astronomical data sets over the
past few years, mainly through large digital sky surveys.  Virtual
Observatory (VO) concept represents a scientific and technological
framework needed to cope with this data flood. We review some of
the applied statistics and computing challenges posed by the
analysis of large and complex data sets expected in the VO-based
research. The challenges are driven both by the size and the
complexity of the data sets (billions of data vectors in parameter
spaces of tens or hundreds of dimensions), by the heterogeneity of
the data and measurement errors, the selection effects and
censored data, and by the intrinsic clustering properties
(functional form, topology) of the data distribution in the
parameter space of observed attributes.
Examples of scientific questions one may wish to address include:
objective determination of the numbers of object classes present
in the data, and the membership probabilities for each source;
searches for unusual, rare, or even new types of objects and
phenomena; discovery of physically interesting multivariate
correlations which may be present in some of the clusters; etc.
%This paper is followed by a commentary by statistician D.\ Cook.
\end{abstract}

\section{Towards a Virtual Observatory}

Observational astronomy is undergoing a paradigm shift.  This
revolutionary change is driven by the enormous technological
advances in telescopes and detectors (e.g., large digital arrays),
the exponential increase in computing capabilities, and the
fundamental changes in the observing strategies used to gather the
data.  In the past, the usual mode of observational astronomy was
that of a single astronomer or small group performing observations
of a small number of objects (from single objects and up to some
hundreds of objects). This is now changing: large digital sky
surveys over a range of wavelengths, from radio to x-rays, from
space and ground are becoming the dominant source of observational
data.  Data-mining of the resulting digital sky archives is
becoming a major venue of the observational astronomy.  The
optimal use of the large ground-based telescopes and space
observatories is now as a follow-up of sources selected from large
sky surveys.  This trend is bound to continue, as the data volumes
and data complexity increase.  The very nature of the
observational astronomy is thus changing rapidly. See, e.g.,
Szalay \& Gray (2001) for a review.

The existing surveys already contain many Terabytes of data, from
which catalogs of many millions, or even billions of objects are
extracted.  For each object, some tens or even hundred parameters
are measured, most (but not all) with quantifiable errors.
Forthcoming projects and sky surveys are expected to deliver data
volumes measured in Petabytes.  For example, a major new area for
exploration will be in the time domain, with a number of ongoing
or forthcoming surveys aiming to map large portions of the sky in
a repeated fashion, down to very faint flux levels. These synoptic
surveys will be generating Petabytes of data, and they will open a
whole new field of searches for variable astronomical objects.

This richness of information is hard to translate into a derived
knowledge and physical understanding.  Questions abound: How do we
explore datasets comprising hundreds of millions or billions of
objects each with dozens of attributes? How do we objectively
classify the detected sources to isolate subpopulations of
astrophysical interest? How do we identify correlations and
anomalies within the data sets? How do we use the data to
constrain astrophysical interpretation, which often involve highly
non-linear parametric functions derived from fields such as
physical cosmology, stellar structure, or atomic physics? How do
we match these complex data sets with equally complex numerical
simulations, and how do we evaluate the performance of such
models?

The key task is now to enable an efficient and complete scientific
exploitation of these enormous data sets.  The problems we face
are inherently statistical in nature.  Similar situations exist in
many other fields of science and applied technology today.  This
poses many technical and conceptual challenges, but it may lead to
a whole new methodology of doing science in the information-rich
era.

In order to cope with this data flood, the astronomical community
started a grassroots initiative, the National (and ultimately
Global) Virtual Observatory (NVO).  The NVO would federate
numerous large digital sky archives, provide the information
infrastructure and standards for ingestion of new data and
surveys, and develop the computational and analysis tools with
which to explore these vast data volumes.  Recognising the urgent
need, the National Academy of Science Astronomy and Astrophysics
Survey Committee, in its new decadal survey {\em Astronomy and
Astrophysics in the New Millennium} (McKee, Taylor, \etal 2001)
recommends, as a first priority, the establishment of a National
Virtual Observatory (NVO).

The NVO would provide new opportunities for scientific discovery
that were unimaginable just a few years ago.  Entirely new and
unexpected scientific results of major significance will emerge
from the combined use of the resulting datasets, science that
would not be possible from such sets used singly.  In the words of
a ``white paper'' \footnote{ Available at
http://www.arXiv.org/abs/astro-ph/0108115, and also published in
Brunner, Djorgovski, \& Szalay (2001), p.~353. } prepared by an
interim steering group the NVO will serve as {\em an engine of
discovery for astronomy.}

Implementation of the NVO involves significant technical
challenges on many fronts, and in particular the {\em data
analysis}. Whereas some of the NVO science would be done in the
image (pixels) domain, and some in the interaction between the
image and catalog domains, it is anticipated that much of the
science (at least initially) will be done purely in the catalog
domain of individual or federated sky surveys.  A typical data set
may be a catalog of $\sim 10^8 - 10^9$ sources with $\sim 10^2$
measured attributes each, i.e., a set of $\sim 10^9$ data vectors
in a $\sim 100$-dimensional parameter space.

Dealing with the analysis of such data sets is obviously an
inherently multivariate statistical problem.  Complications
abound: parameter correlations will exist; observational limits
(selection effects) will generally have a complex geometry; for
some of the sources some of the measured parameters may be only
upper or lower limits; the measurement errors may vary widely;
some of the parameters will be continuous, and some discrete, or
even without a well-defined metric; etc.  In other words, analysis
of the NVO data sets will present many challenging problems for
multivariate statistics, and the resulting astronomical
conclusions will be strongly affected by the correct application
of statistical tools.

We review some important statistical challenges raised by the NVO.
These include the classification and extraction of desired
subpopulations, understanding the relationships between observed
properties within these subpopulations, and linking the
astronomical data to astrophysical models. This may require a
generation of new methods in data mining, multivariate clustering
and analysis, nonparametric and semiparametric estimation and
model and hypothesis testing.

\section{Clustering analysis challenges in a VO}

The exploration of observable parameter spaces, created by
combining of large sky surveys over a range of wavelengths, will
be one of the chief scientific purposes of a VO. This includes an
exciting possibility of discovering some previously unknown types
of astronomical objects or phenomena (see Djorgovski \etal 2001a,
2001b, 2001c for reviews).

A complete observable parameter space axes include quantities such
as the object coordinates, velocities or redshifts, sometimes
proper motions, fluxes at a range of wavelength (i.e., spectra;
imaging in a set of bandpasses can be considered a form of a very
low resolution spectroscopy), surface brightness and image
morphological parameters for resolved  sources, variability (or,
more broadly, power spectra) over a range of time scales, etc.~
Any given sky survey samples only a small portion of this grand
observable parameter space, and is subject to its own selection
and measurement limits, e.g., limiting fluxes, surface brightness,
angular resolution, spectroscopic resolution, sampling and
baseline for variability if multiple epoch observations are
obtained, etc.

A major exploration technique envisioned for the NVO will be
unsupervised clustering of data vectors in some parameter space of
observed properties of detected sources.  Aside from the
computational challenges with large numbers of data vectors and a
large dimensionality, this poses some highly non-trivial
statistical problems.  The problems are driven not just by the
$size$ of the data sets, but mainly  (in the statistical context)
by the {\it heterogeneity and intrinsic complexity of the data}.

A typical VO data set may consist of $\sim 10^9$ data vectors in
$\sim 10^2$ dimensions.   These are measured source attributes,
including positions, fluxes in different bandpasses, morphology
quantified through different moments of light distribution and
other suitably constructed parameters, etc.  Some of the
parameters would be primary measurements, and others may be
derived attributes, such as the star-galaxy classification, some
may be ``flags'' rather than numbers, some would have error-bars
associated with them, and some would not, and the error-bars may
be functions of some of the parameters, e.g., fluxes. Some
measurements would be present only as upper or lower limits. Some
would be affected by ``glitches'' due to instrumental problems,
and if a data set consists of a merger of two or more surveys,
e.g., cross-matched optical, infrared, and radio (and this would
be a common scenario within a VO), then some sources would be
misidentified, and thus represent erroneous combinations of
subsets of data dimensions.  Surveys would be also affected by
selection effects operating explicitly on some parameters (e.g.,
coordinate ranges, flux limits, etc.), but also mapping onto some
other data dimensions through correlations of these properties;
some selection effects may be unknown.

Physically, the data set may consist of a number of distinct
classes of objects, such as stars (including a range of spectral
types), galaxies (including a range of Hubble types or
morphologies), quasars, etc. Within each object class or subclass,
some of the physical properties may be correlated, and some of
these correlations may be already known and some as yet unknown,
and their discovery would be an important scientific result by
itself. Some of the correlations may be spurious (e.g., driven by
sample selection effects), or simply uninteresting (e.g., objects
brighter in one optical bandpass will tend to be brighter in
another optical bandpass). Correlations of independently measured
physical parameters represent a reduction of the statistical
dimensionality in a multidimensional data parameter space, and
their discovery may be an integral part of the clustering
analysis.

Typical scientific questions posed may be:

\begin{itemize}

\item How many statistically distinct classes of objects are in this
data set, and which objects are to be assigned to which class, along
with association probabilities?

\item Are there any previously unknown classes of objects, i.e.,
statistically significant ``clouds'' in the parameter space
distinct from the ``common'' types of objects (e.g., normal stars
or galaxies)? An application may be separating quasars from
otherwise morphologically indistinguishable normal stars.

\item Are there rare outliers, i.e., individual objects with a low
probability of belonging to any one of the dominant classes?  Examples
may include known, bur relatively rare types of objects such as
high-redshift quasars, brown dwarfs, etc., but also previously unknown
types of objects; finding any such would be a significant discovery.

\item Are there interesting (in general, multivariate) correlations
among the properties of objects in any given class, and what are the
optimal analytical expressions of such correlations?  An example may be
the ``Fundamental Plane'' of elliptical galaxies, a set of bivariate
correlations obeyed by this Hubble type, but no other types of galaxies
(see, e.g., Djorgovski 1992, 1993, and Djorgovski \etal 1995, for reviews).

\end{itemize}

The complications include the following:

\begin{enumerate}

\item Construction of these complex data sets, especially if multiple sky
      surveys, catalogs, or archives are being federated (an essential VO
      activity) will inevitably be imperfect, posing quality control problems
      which must be discovered and solved first, before the scientific
      exploration starts.  Sources may be mismatched, there will be some gross
      errors or instrumental glitches within the data, subtle systematic
      calibration errors may affect pieces of the large data sets, etc.

\item The object classes form multivariate ``clouds'' in the parameter
      space, but these clouds in general need not be Gaussian:  some may have
      a power-law or exponential tails in some or all of the dimensions, and
      some may have sharp cutoffs, etc.

\item The clouds may be well separated in some of the dimensions, but not
      in others.  How can we objectively decide which dimensions are
      irrelevant, and which ones are useful?

\item The {\it topology} of clustering may not be simple:  there may be
      clusters within clusters, holes in the data distribution (negative
      clusters?), multiply-connected clusters, etc.

\item All of this has to take into the account the heterogeneity of
      measurements, censored data, incompleteness, etc.

\end{enumerate}

The majority of the technical and methodological challenges in
this quest derive from the expected heterogeneity and intrinsic
complexity of the data, including treatment of upper an lower
limits, missing data, selection effects and data censoring, etc.
These issues affect the proper statistical description of the
data, which then must be reflected in the clustering algorithms.

Related to this are the problems arising from the data modeling.
The commonly used mixture-modeling assumption of clusters
represented as multivariate Gaussian clouds is rarely a good
descriptor of the reality. Clusters may have non-Gaussian shapes,
e.g., exponential or power-law tails, asymmetries, sharp cutoffs,
etc.~ This becomes a critical issue in evaluating the membership
probabilities in partly overlapping clusters, or in a search for
outliers (anomalous events) in the tails of the distributions.  In
general, the proper functional forms for the modeling of clusters
are not known {\it a priori}, and must be discovered from the
data.  Applications of non-parametric techniques may be essential
here. A related, very interesting problem is posed by the {\it
topology} of clustering, with a possibility of multiply-connected
clusters or gaps in the data (i.e., negative clusters embedded
within the positive ones), hierarchical or multi-scale clustering
(i.e., clusters embedded within the clusters) etc.

The clusters may be well separated in some of the dimensions, but
not in others. How can we objectively decide which dimensions are
irrelevant, and which ones are useful?  An automated and objective
rejection of the ``useless'' dimensions, perhaps through some
statistically defined entropy criterion, could greatly simplify
and speed up the clustering analysis.

Once the data are partitioned into distinct clusters, their
analysis and interpretation starts.  One question is, are there
interesting (in general, multivariate) correlations among the
properties of objects in any given cluster?  Such correlations may
reflect interesting new astrophysics (e.g.,, the stellar main
sequence, the Tully-Fisher and Fundamental Plane correlations for
galaxies, etc.), but at the same time complicate the statistical
interpretation of the clustering.  They would be in general
restricted to a subset of the dimensions, and not present in the
others.  How do we identify all of the interesting correlations,
and discriminate against the ``uninteresting'' observables?

Here we describe some of our experiments to date, and outline some possible
avenues for future exploration.

\section{Examples and some possible approaches}

Separation of the data into different types of objects, be it
known or unknown in nature, can be approached as a problem in
automated classification or clustering analysis.  This is a part
of a more general and rapidly growing field of Data Mining (DM)
and Knowledge Discovery in Databases (KDD).  We see here great
opportunities for collaborations between astronomers and computer
scientists and statisticians.  For an overview of some of the
issues and methods, see, e.g., Fayyad \etal (1996b) .

If applied in the catalog domain, the data can be viewed as a set
of $n$ points or vectors in an $m$-dimensional parameter space,
where $n$ can be in the range of many millions or even billions,
and $m$ in the range of a few tens to hundreds.  The data may be
clustered in $k$ statistically distinct classes, which could be
modeled, e.g., as multivariate Gaussian clouds, and which hopefully
correspond to physically distinct classes of objects (e.g., stars,
galaxies, quasars, etc.).  This is schematically illustrated in
Figure 1.

\begin{figure}[b]
\begin{center}
\includegraphics[width=.8\textwidth]{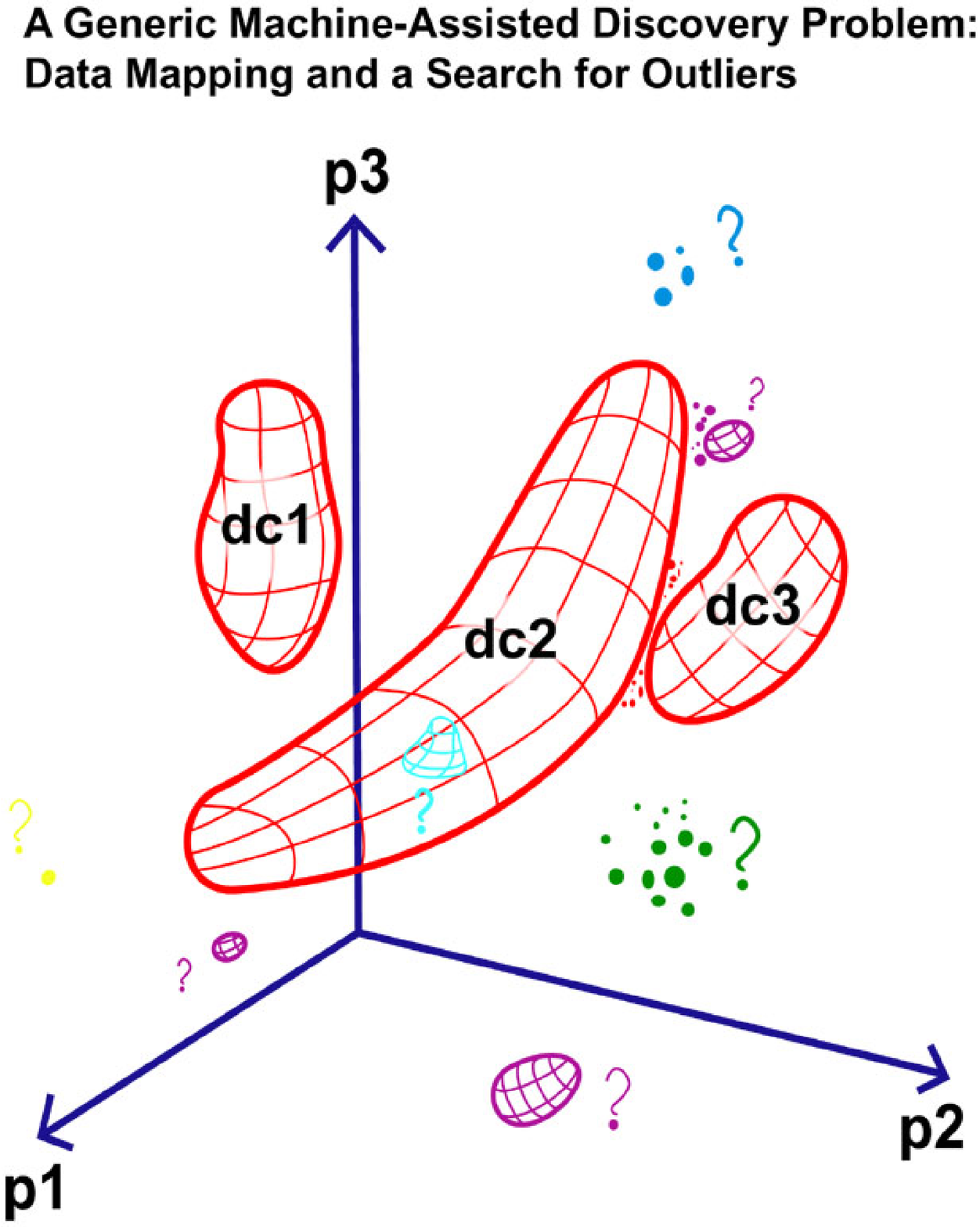}
\end{center}
\caption[]{
A schematic illustration of the problem of clustering
analysis in some parameter space.  In this example, there are 3
dimensions, $p1$, $p2$, and $p3$ (e.g., some flux ratios or
morphological parameters), and most of the data points belong to 3
major clusters, denoted $dc1$, $dc2$, and $dc3$ (e.g., stars,
galaxies, and ordinary quasars).  One approach is to isolate these
major classes of objects for some statistical studies, e.g., stars
as probes of the Galactic structure, or galaxies as probes  of the
large scale structure of the universe, and filter out the
``anomalous'' objects.  A complementary view is to look for other,
less populated, but statistically significant, distinct clusters
of data points, or even individual outliers, as possible examples
of rare or unknown types of objects.  Another possibility is to
look for holes (negative clusters) within the major clusters, as
they may point to some interesting physical phenomenon -- or to a
problem with the data.
}
\label{eps1}
\end{figure}

If the number of object classes $k$ is known (or declared) {\it a
priori}, and training data set of representative objects is
available, the problem reduces to supervised classification, where
tools such as Artificial Neural Nets or Decision Trees can be
used.  This is now commonly done for star-galaxy separation in sky
surveys (e.g., Odewahn \etal 1992, or Weir \etal 1995). Searches for
known types of objects with predictable signatures in the
parameter space (e.g., high-$z$ quasars) can be also cast in this
way.

However, a more interesting and less biased approach is where the
number of classes $k$ is not known, and it has to be derived from
the data themselves. The problem of unsupervised classification is
to determine this number in some objective and statistically sound
manner, and then to associate class membership probabilities for
all objects.  Majority of objects may fall into a small number of
classes, e.g., normal stars or galaxies.  What is of special
interest are objects which belong to much less populated clusters,
or even individual outliers with low membership probabilities for
any major class. Some initial experiments with unsupervised
clustering algorithms in the astronomical context include, e.g.,
Goebel \etal (1989), Weir \etal (1995), de Carvalho \etal (1995),
and Yoo \etal (1996), but a full-scale application to major
digital sky surveys yet remains to be done. Intriguing
applications which addressed the issue of how many statistically
distinct classes of GRBs are there (Mukherjee \etal 1998, Rogier
\etal 2000, Hakkila \etal 2000).

In many situations, scientifically informed input is needed in
designing the clustering experiments.  Some observed parameters
may have a highly significant, large dynamical range, dominate the
sample variance, and naturally invite division into clusters along
the corresponding parameter axes; yet they may be completely
irrelevant or uninteresting scientifically.  For example, if one
wishes to classify sources of the basic of their broad-band
spectral energy distributions (or to search for objects with
unusual spectra), the mean flux itself is not important, as it
mainly reflects the distance; coordinates on the sky may be
unimportant (unless one specifically looks for a spatial
clustering); etc.  Thus, a clustering algorithm may divide the
data set along one or more of such axes, and completely miss the
really scientifically interesting partitions, e.g., according to
the colors of objects.

One method we have been experimenting with (applied on the various
data sets derived from DPOSS) is the Expectation Maximisation (EM)
technique, with the Monte Carlo Cross Validation (MCCV) as the way
of determining the maximum likelihood number of the clusters.

This may be a computationally very expensive problem. For the
simple $K$-means algorithm, the computing cost scales as $K
~\times ~N ~\times ~I ~\times ~D$, where $K$ is the number of
clusters chosen {\it a priori}, $N$ is the number of data vectors
(detected objects), $I$ is the number of iterations, and $D$ is
the number of data dimensions (measured parameters per object).
For the more powerful Expectation Maximisation technique, the cost
scales as $K ~\times ~N ~\times ~I ~\times ~D^2$, and again one
must decide {\it a priori} on the value of $K$.  If this number
has to be determined intrinsically from the data, e.g., with the
Monte Carlo Cross Validation method, the cost scales as $M ~\times
~K_{max}^2 ~\times ~N ~\times ~I ~\times ~D^2$ where $M$ is the
number of Monte Carlo trials/partitions, and $K_{max}$ is the
maximum number of clusters tried. Even with the typical numbers
for the existing large digital sky surveys ($N \sim 10^8 - 10^9$,
$D \sim 10 - 100$) this is already reaching in the realm of
Terascale computing, especially in the context of an interactive
and iterative application of these analysis tools.  Development of
faster and smarter algorithms is clearly a priority.

One technique which can simplify the problem is the
multi-resolution clustering.  In this regime, expensive parameters
to estimate, such as the number of classes and the initial broad
clustering are quickly estimated using traditional techniques, and
then one could proceed to refine the model locally and globally by
iterating until some objective statistical (e.g., Bayesian)
criterion is satisfied.

One can also use intelligent sampling methods where one forms
``prototypes''of the case vectors and thus reduces the number of
cases to process.  Prototypes can be determined from simple
algorithms to get a rough estimate, and then refined using more
sophisticated techniques.  A clustering algorithm can operate in
prototype space.  The clusters found can later refined by locally
replacing each prototype by its constituent population and
reanalyzing the cluster.

Techniques for dimensionality reduction, including principal
component analysis and others can be used as preprocessing
techniques to automatically derive the dimensions that contain
most of the relevant information.

\section{Concluding comments}

Given this computational and statistical complexity, blind
applications of the commonly used (commercial or home-brewed)
clustering algorithms could produce some seriously misleading or
simply wrong results.  The clustering methodology must be robust
enough to cope with these problems, and the outcome of the
analysis must have a solid statistical foundation.

In our experience, design and application of clustering algorithms
must involve close, working collaboration between astronomers and
computer scientists and statisticians.  There are too many
unspoken assumptions, historical background knowledge specific to
the given discipline, and opaque jargon; constant communication
and interchange of ideas are essential.

The entire issue of discovery and interpretation of multivariate
correlations in these massive data sets has not really been
addressed so far.  Such correlations may contain essential clues
about the physics and the origins of various types of astronomical
objects.

Effective and powerful data visualization, applied in the
parameter space itself, is another essential part of the
interactive clustering analysis. Good visualisation tools are also
critical for the interpretation of results, especially in an
iterative environment.  While clustering algorithms can assist in
the partitioning of the data space, and can draw the attention to
anomalous objects, ultimately a scientist guides the experiment
and draws the conclusions. It is very hard for a human mind to
really visualise clustering or correlations in more than a few
dimensions, and yet both interesting clusters and multivariate
correlations with statistical dimensionality $> 10$ or even higher
are likely to exits, and possibly lead to some crucial new
astrophysical insights.  Perhaps the right approach would be to
have a good visualisation embedded as a part of an interactive and
iterative clustering analysis.

Another key issue is interoperability and reusability of
algorithms and models in a wide variety of problems posed by a
rich data environment such as federated digital sky surveys in a
VO.  Implementation of clustering analysis algorithms must be done
with this in mind.

Finally, scientific verification and evaluation, testing, and
follow-up on any of the newly discovered classes of objects,
physical clusters discovered by these methods, and other
astrophysical analysis of the results is essential in order to
demonstrate the actual usefulness of these techniques for a VO or
other applications.  Clustering analysis can be seen as a prelude
to the more traditional type of astronomical studies, as a way of
selecting of interesting objects of samples, and hopefully it can
lead to advances in statistics and applied computer science as
well.

\section{Acknowledgments}
We wish to thank numerous collaborators, including R. Gal, S.
Odewahn, R. de Carvalho, T. Prince, J. Jacob, D. Curkendall, and
many others. This work was supported in part by the NASA grant
NAG5-9482, and by private foundations.  Finally, we thank the
organizers for a pleasant and productive meeting.

\section{References}

\pp Boller, T., Meurs, E., \& Adorf, H.-M. 1992, A\&A, 259, 101

\pp Brunner, R.J., Djorgovski, S.G., \& Szalay, A.S. (editors) 2001a,
    {\sl Virtual Observatories of the Future}, ASPCS vol. 221.

\pp Brunner, R., Djorgovski, S.G., Gal, R.R., Mahabal, A., \& Odewahn, S.C.
    2001b, in: {\sl Virtual Observatories of the Future}, eds. R.~Brunner,
    S.G.~Djorgovski \& A.~Szalay, ASPCS, 225, 64

\pp Burl, M., Asker, L., Smyth, P., Fayyad, U., Perona, P., Crumpler, L., \&
    Aubelle, J. 1998, Mach.~Learning, 30, 165

\pp de Carvalho, R., Djorgovski, S., Weir, N., Fayyad, U., Cherkauer, K.,
    Roden, J., \& Gray, A. 1995, in {\sl Astronomical Data Analysis
    Software and Systems IV}, eds. R. Shaw \etal, ASPCS, 77, 272

\pp Djorgovski, S.G. 1992,
    in: {\sl Cosmology and Large-Scale Structure in the Universe},
    ed. R. de Carvalho, ASPCS, 24, 19

\pp Djorgovski, S.G. 1993,
    in: {\sl The Globular Cluster -- Galaxy Connection},
    eds. G.~Smith \& J.~Brodie, ASPCS, 48, 496

\pp Djorgovski, S.G., Pahre, M.A., \& de Carvalho, R.R. 1995,
    in: {\sl Fresh Views of Elliptical Galaxies},
    eds. A.~Buzzoni \etal, ASPCS, 86, 129

\pp Djorgovski, S.G., Mahabal, A., Brunner, R., Gal, R.R., Castro, S.,
    de Carvalho, R.R., \& Odewahn, S.C.  2001a,
    in: {\sl Virtual Observatories of the Future},
    eds. R.~Brunner, S.G.~Djorgovski \& A.~Szalay, ASPCS, 225, 52
    [astro-ph/0012453]

\pp Djorgovski, S.G., Brunner, R., Mahabal, A., Odewahn, S.C.,
    de Carvalho, R.R., Gal, R.R., Stolorz, P., Granat, R., Curkendall, D.,
    Jacob, J., \& Castro, S.  2001b,
    in: {\sl Mining the Sky}, eds. A.J.~Banday \etal,
    ESO Astrophysics Symposia, Berlin: Springer Verlag, p.~305
    [astro-ph/0012489]

\pp Djorgovski, S.G., Mahabal, A., Brunner, R., Williams, R., Granat, R.,
    Curkendall, D., Jacob, J., \& Stolorz, P. 2001c,
    in: {\sl Astronomical Data Analysis}, eds. J.-L. Starck \& F. Murtagh,
    {\sl Proc. SPIE} {\bf 4477}, in press
    [astro-ph/0108346]

\pp Fayyad, U., Djorgovski, S.G., \& Weir, W.N. 1996a, in:
    {\sl Advances in Knowledge Discovery and Data Mining},
    eds. U. Fayyad \etal, Boston: AAAI/MIT Press, p.~471

\pp Fayyad, U., Piatetsky-Shapiro, G., Smyth, P., \& Uthurusamy, R.
    (eds.) 1996b, {\sl Advances in Knowledge Discovery and Data Mining},
    Boston: AAAI/MIT Press

\pp Goebel, J., Volk, K., Walker, H., Gerbault, F., Cheeseman, P., Self, M.,
    Stutz, J., \& Taylor, W. 1989, A\&A, 222, L5

\pp Hakkila, J., Haglin, D., Pendleton, G., Mallozzi, R., Meegan, C., \&
    Rogier, R. 2000, ApJ, 538, 165

\pp Mukherjee, S., Feigelson, E., Babu, J., Murtagh, F., Fraley, C., \&
    Raftery, A. 1998, ApJ, 508, 314

\pp Odewahn, S.C., Stockwell, E., Pennington, R., Humphreys, R., \& Zumach, W.
    1992, AJ, 103, 318

\pp Paczy\'nski, B. 2000, PASP, 112, 1281

\pp Rogier, R., Hakkila, J., Haglin, D., Pendleton, G., \& Mallozzi, R. 2000,
    in: {\sl Gamma-Ray Bursts, 5th Hunsville Symp.}, eds. R. Kippen \etal,
    AIP Conf. Proc. 526, 38

\pp Szalay, A., \& Gray, J. 2001, Science, 293, 2037

\pp Weir, N., Fayyad, U., \& Djorgovski, S. 1995, AJ, 109, 2401

\pp Yoo, J., Gray, A., Roden, J., Fayyad, U., de Carvalho, R., \& Djorgovski, S.
    1996, in: {\sl Astronomical Data Analysis Software and Systems V},
    eds. G. Jacoby \& J. Barnes, ASPCS, 101, 41

\bigskip

\noindent
{\sl Note added in the preprint version of the paper:}~
Interested reader may find a lot of information about the VO concept,
and some useful links at the NVO Science Definition Team website,
http://www.nvosdt.org

\end{document}